\documentclass[twocolumn,aps,pra,superscriptaddress,showpacs,tightenlines]{revtex4}
\usepackage{amsmath}
\usepackage{amsfonts}
\usepackage{graphicx}
\usepackage{epsfig}
\usepackage{color}
\usepackage[colorlinks,citecolor=blue]{hyperref}
\begin{document}

\title{Single-photon-triggered quantum chaos}
\author{Gui-Lei Zhu}
\affiliation{School of physics, Huazhong University of Science and Technology, Wuhan 430074, China}

\author{Xin-You L\"{u}}
\email{xinyoulu@hust.edu.cn}
\affiliation{School of physics, Huazhong University of Science and Technology, Wuhan 430074, China}

\author{Li-Li Zheng}
\affiliation{School of physics, Huazhong University of Science and Technology, Wuhan 430074, China}

\author{Zhi-Ming Zhan}
\affiliation{School of Physics and Information Engineering, Key Laboratory of Optoelectronic Chemical Materials and Devices of Ministry of Education, Jianghan University, Wuhan, 430074, China}

\author{Franco Nori}
\affiliation{Theoretical Quantum Physics Laboratory, RIKEN Cluster for Pioneering Research, Wako-shi, Saitama 351-0198, Japan}
\affiliation{Physics Department, The university of Michigan, Ann Arbor, Michgan 48109-1040, USA}

\author{Ying Wu}
\email{yingwu2@126.com}
\affiliation{School of physics, Huazhong University of Science and Technology, Wuhan 430074, China}
\date{\today}
\begin{abstract}

{\color{black}We demonstrate how to manipulate quantum chaos with a single photon in a hybrid quantum device combining cavity QED and optomechanics. Specifically, we show that this system changes between integrable and chaotic relying on the photon-state of the injected field. This onset of chaos originates from the photon-dependent chaotic threshold of the qubit-field coupling induced by the optomechanical interaction. By deriving the Loschmidt Echo we observe clear differences in the sensitivity to perturbations in the regular versus chaotic regimes. We also present classical analog of this chaotic behavior, and find good correspondence between chaotic quantum dynamics and classical physics.
Our work opens up a new route to achieve quantum manipulations, which are crucial elements in engineering new types of on-chip quantum devices and quantum information science.}
\end{abstract}
\maketitle
\section{Introduction}\label{sec1}
{\color{black} Chaos plays an important role in
most fields of classical physics~\cite{Hilborn2000} and quantum mechanics~\cite{Gutzwiller1990}. The study of chaos is not only significant from the perspective of understanding fundamental physics, but also for potential applications in achieving secure communication\,\cite{Van1998,Argyris2005,Colet1994}, enhancing tunneling rates\,\cite{Lin1990,Steck2001,Dembowski2000}, and building operable quantum computers\,\cite{Georgeot2001}. The field of quantum chaos, studying how classical chaotic dynamics manifests itself in quantum mechanics, has achieved spectacular advances in recent years~\cite{Lemos2012,Albert2011,Neill,Xiao2017,Gao2015,Franco2015}. However, previous works on quantum chaos merely focus on certain simple quantum systems~\cite{Dicke1954,Emary2003,Perez2011,Shammah2018,Chaudhury2009}, e.g., Dicke model, kicked top, nuclear model, etc.

Cavity optomechanics studies the quantum effects induced by the radiation-pressure interaction between the electromagnetic and mechanical systems\,\cite{reviews,Wu2015}, which provides a platform for manipulating the bosonic field. Especially, the quadratic optomechanical coupling\,\cite{Lu2015,Thompson2008,Sankey2010,Nunnenkamp2010,Xuereb2013,Zhu2018,Kim2015,Jie-Qiao} has attracted much attention even though it is very weak. Recently, the development of nano-fabricated optomechanical technologies makes it possible to introduce the optomechanical interaction into cavity QED or other systems\,\cite{Franco2018,Qinwei,Clerk2018,Ze-Liang,Treutlein2014,Bin2019,Zheng2019}. {\color{black}Here we propose a possible scheme in a cavity optomechanical system ``doped" with an atomic ensemble realizing a Dicke model (DM) to study its chaotic behavior.} Such proposal has potential applications for inspiring various on-chip quantum devices.

Here we study how to manipulate quantum chaos with a single photon in a hybrid quantum model by combining an optomechanical system with cavity QED.
The proposed system can be changed between its quasi-integrable regime and its quantum chaotic one by preparing different photon states of the ancillary mode. In this hybrid model, the corresponding chaotic threshold of qubit-field coupling could be very weak.
Physically, the introduced quadratic optomechanical coupling can effectively change the qubit-field interaction strength, depending on the photon number of the ancillary cavity. Thus, the hybrid model studied here has a {\it photon-dependent chaotic threshold} in its qubit-field coupling, which ultimately leads to single-photon-triggered quantum chaos. Apart from the fundamental interest in exploring the quantum-classical correspondence\,\cite{Zurek}, our work will inspire further investigations regarding photon-dependent quantum dynamic effects in hybrid cavity QED. Moreover, this work, combining  quantum chaos with single-photon technologies, could be used for chaos-assisted communications\,\cite{Argyris2008,Nguimdo2011} and various single-photon devices\,\cite{Hadfield2009,zhou2013,Baur2014}.}

\section{Model} \label{sec2}

\begin{figure}
\includegraphics[width=7.5cm]{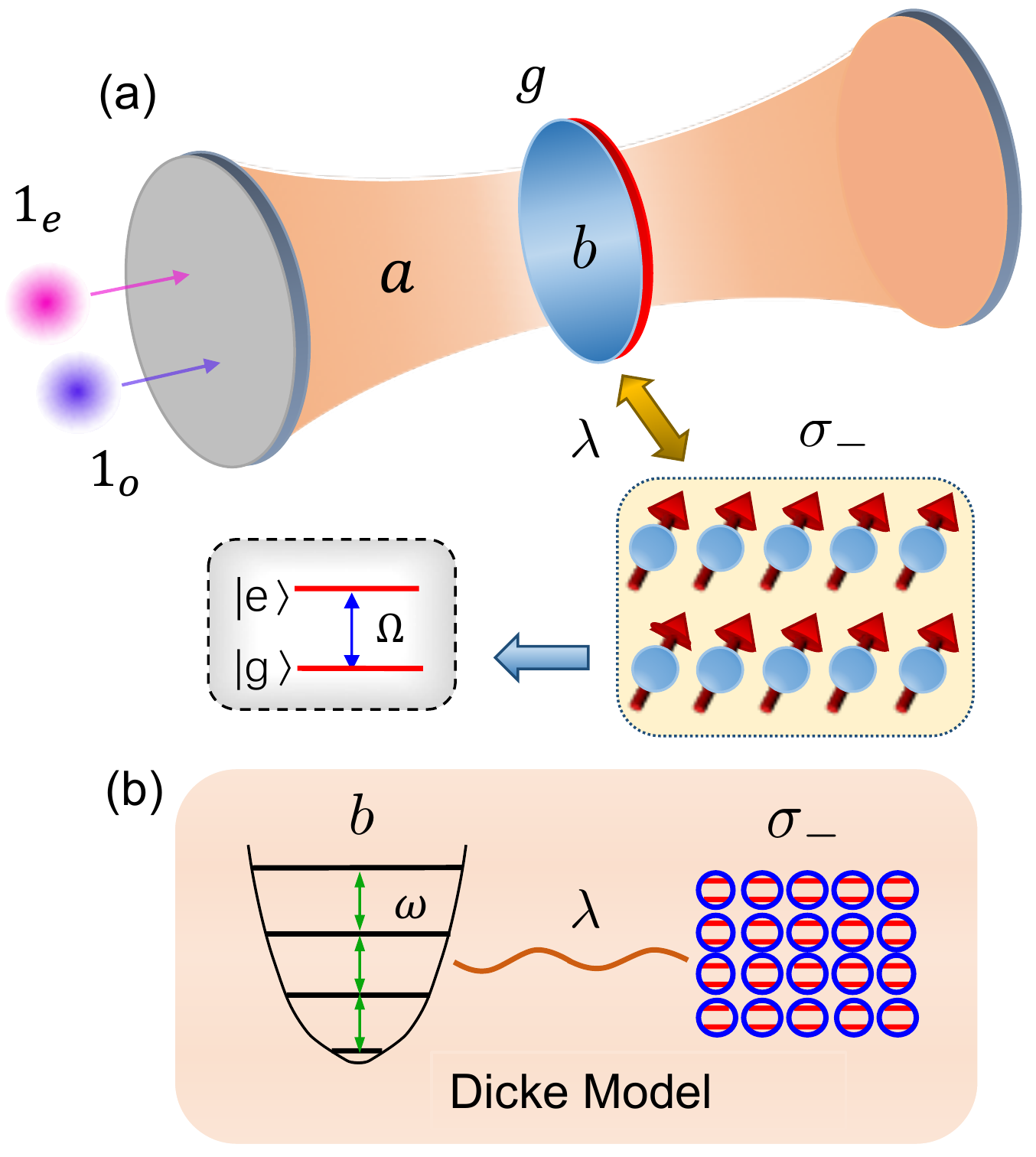}
\caption{(a) Schematic diagram of the physical implementation of the generalized hybrid model containing a normal Dicke model (DM) and an ancillary cavity. The ancillary cavity $a_l  (l=e,o)$ couples quadratically with the normal DM (a bosonic single-mode $b$ interacts with $N$ two-level systems $\sigma_-$ with dipole coupling strength $\lambda$), with optomechanical coupling strengths $g_e=g$ and $g_o=ge^{i\pi}$. (b) The Dicke model could be implemented by coupling a mechanical resonator to a collection of two-level systems.}
\label{fig1}
\end{figure}

As depicted in Fig.\,\ref{fig1}(a), we consider a hybrid optomechanical system consisting of a normal Dicke model and an ancillary cavity. The total Hamiltonian is given by ($\hbar=1$),
\begin{align}
H=H_{\rm an}+H_{\rm dm}-\sum_{l=e,o}{g_{l}}a_l^{\dagger}a_l(b^{\dagger}+b)^2,\label{H_or}
\end{align}
where $a_l$ ($a_l^{\dagger}$) and $b$ ($b^{\dagger}$) are the annihilation (creation) operators of the ancillary mode and single-mode bosonic field of the DM, respectively. Note that $g_{l}$ $(l=e,o)$ quantifies the optomechanical quadratic coupling strength between the two bosonic modes $a_l$ and $b$, where $g_e=g$ and $g_o=ge^{i\pi}$ manifest that the bosonic mode $b$ interacts with $a_l$ at the position of even or odd number of half wavelengths of the cavity\,\cite{Bhattacharya2008}. Here $H_{\rm an}=\sum_{l=e,o}\omega_l a_l^{\dagger}a_l$, and $\omega_l$ is the frequency of the ancillary cavity mode. The DM Hamiltonian reads: $H_{\rm dm}=\omega b^{\dagger}b+\Omega J_z+\frac{\lambda}{\sqrt{N}}(b^{\dagger}+b)J_x,$
with a collective qubit-field coupling strength $\lambda$. For the $N$ qubits, $\Omega$ is the excitation energy; meanwhile, the collective pseudospin operators obey the $SU_{2}$ algebra, and they are given by $J_{z}=(1/2)\sum^N_{i=1}\sigma_z$, $J_{\pm}=\sum^N_{i=1}\sigma_\pm$, and $J_x=J_-+J_+$. Due to parity conservation, $\left[H_{\rm tot},\Pi\right]=0$, the Hilbert space of $H_{\rm tot}$ can be separated into two noninteracting parts. Here we note that $\Pi=e^{i\pi\mathcal{N}}$, and $\mathcal{N}=b^{\dagger}b+J_z+N/2$ is the total excitation number of the system (excluding the ancillary modes $a$).

It should be stressed that in the Hamiltonian $H$, the quadratic coupling term produces a photon-dependent modification on the potential of field $b$. We consider the ancillary mode to be prepared in the Fock state $|n_e \,n_o\rangle_a(n_e=n_o+n$, and $n_e,n_o,n=0,1,2...)$. Then the photon number operator $a_l^{\dagger}a_l$ could be replaced by an algebraic number $n_l$. More specifically, provided that the initial ancillary mode is in $|00\rangle_a (n=0)$, injecting a photon ($n_e=1$) into the cavity, then the ancillary mode would be in $|10\rangle_a$ $(n=1)$. However, if another photon ($n_o=1$) is subsequently injected into the cavity, then the ancillary mode is in $|11\rangle_a$ $(n=0)$, which is physically equivalent to $|00\rangle_a$. In other words, injecting two different photons ($n_e=1$, $n_o=1$) makes the ancillary mode return to the equivalent initial state, i.e., $|00\rangle_a\rightarrow|10\rangle_a\rightarrow|11\rangle_a$. Hereafter, we only consider the case of $n=0,1$, where $n=0$ corresponds to a normal DM. We now apply a squeezing transformation $b=\cosh(r_{n})b_{n}+\sinh(r_{n})b_{n}^{\dagger}$ with $r_{n}=(-1/4)\ln[1-4ng_l/\omega]$, then system Hamiltonian becomes Dicke-like,
\begin{align}
H_{n}=\Omega J_{z}+\omega_n b_{n}^{\dagger}b_n+\frac{\lambda_n}{\sqrt{N}}(b_{n}^{\dagger}+b_n)J_{x}+C_n, \label{H_n1}
\end{align}
where $\omega_n={\rm exp}(-2r_n)\omega$, $\lambda_n={\rm exp}(r_n)\lambda$, and $C_n=\sum_{l=e,o}n_l\omega_l+[{\rm exp}(-2r_n)-1](\omega/2)$, are the photon-dependent system parameters.
Equation (\ref{H_n1}) shows that the qubit-field coupling strength could be enhanced significantly by adjusting the photon-dependent squeezing parameter $r_n$\,\cite{Lu2015}, which allows the occurrence of single-photon-triggered quantum chaotic behavior.
\section{Single-photon-triggered quantum chaos}

To investigate the quantum chaotic behavior triggered by the single photon of the proposed system, in the following, we present three signatures of quantum chaos, i.e., (1) nearest-neighbor level spacing, (2) Loschmidt Echo, and (3) Poincar\'{e} sections.

\begin{figure}
\centerline{\includegraphics[width=9.0cm]{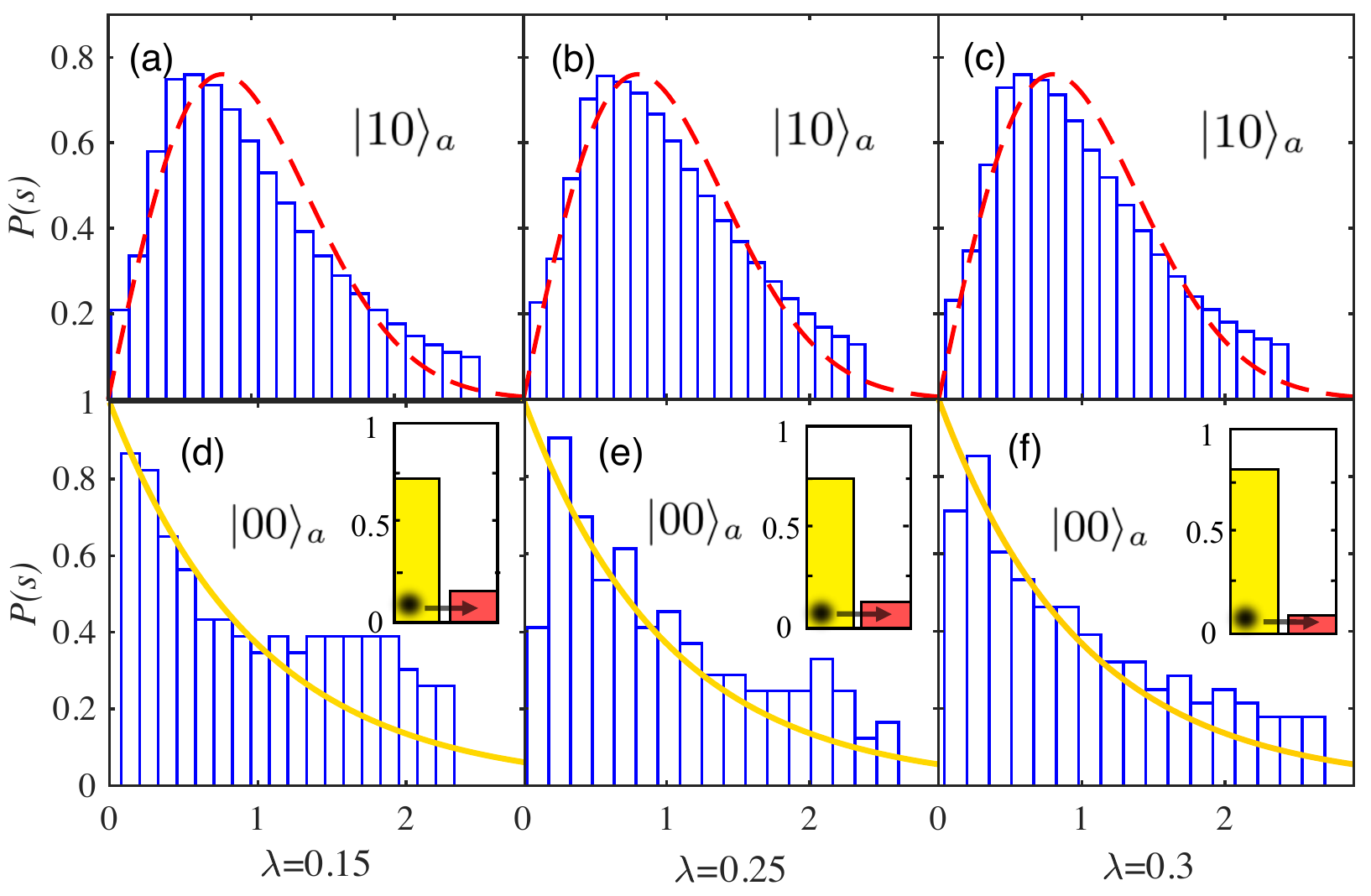}}
\caption{Nearest-neighbor level spacing distribution $P(s)$ for several values of $\lambda$ considering the ancillary mode to be prepared in $|10\rangle_a$ (top) and $|00\rangle_a$ (bottom), respectively. Note that $|10\rangle_a$ corresponds to $n=1$, and $|00\rangle_a$ to $n=0$. The yellow-solid curve is the universal Poissonian distribution and the red-dashed curve shows the Wigner-Dyson distribution. The inserted bar graphs quantitatively present the quantity $\eta$. The yellow and red bars show the case of $n=0$ and $n=1$, respectively. The horizontal black arrow indicates the single-photon triggered quantum chaos. The system parameters are chosen as $\Omega=\omega=1,g=0.23\omega$, and $N=20$.}
\label{fig2}
\end{figure}

\subsection {Level spacing} Quantum chaos manifests itself by the statistical properties of the energy levels Specifically, the character of the energy spectra can be quantified by the nearest-neighbor level-spacing distribution $P(s)$, {\color{black}which has been experimentally verified in Rydberg excitons\,\cite{ABmann2016,Nori} and acoustic resonator\,\cite{Neicu2001}}. The comparison between $P(s)$ and the result obtained by random matrix theory (RMT) produces a signature of quantum chaos~\cite{Bohigas1984,Pechukas1983}. The levels of classical integrable systems are uncorrelated and show a Poissonian statistics, i.e., $P_{\rm p}(s)={\rm exp}(-s)$, indicating strong level clustering\,\cite{Berry}. Conversely, in chaotic systems, the energy level-spacing distribution is well approximated by the Wigner-Dyson function, i.e., $P_{\rm w}(s)=\pi s/2\, {\rm exp}(-\pi s^2/4)$, indicating the eigenvalues of random matrices\,\cite{Berry}.

To investigate the level statistics of the system, we first numerically diagonalize the Hamiltonian $H_n$. Then we apply a general unfolding process; finally the nearest-neighbor level-spacing distribution $P(s)$ is constructed. Figure \,\ref{fig2} shows the $P(s)$ distributions of the proposed system for different values of the qubit-field coupling strength $\lambda$. In the region considered, when the ancillary cavity is prepared in $|00\rangle_a$ [this is a normal DM corresponding to $n=0$, shown in Fig.\,\ref{fig2}(d)-(f)], the spectral statistical property closely follows a Poissonian distribution, which is the counterpart of a classical integrable system. In contrast, it yields the Wigner distribution when the mode $a$ is in $|10\rangle_a$ [shown in Fig.\,\ref{fig2}(a)-(c)], which manifests the chaotic property of the system. The transition of the level statistics from a Poisson-like distribution to a Wigner-like form provides a clear indication that quantum chaos is triggered by a single photon.

To obtain a more quantitative description, we calculate the quantity $\eta =|\int_{0}^{s_0}[P(s)-P_{\rm w}(s)]ds/\int_{0}^{s_0}[P_{\rm p}(s)-P_{\rm w}(s)]ds|$, where $s_0=0.472913...$\,\cite{Georgeot1998}. This $\eta$ characterizes the degree of similarity between $P(s)$ and the normal Poissonian distribution $P_{\rm p}(s)$. Theoretically, $\eta=0$ if $P(s)$ follows the Wigner distribution, while $\eta=1$ for a Poissonian distribution. The inserted graph bars in Fig.\,\ref{fig2} show the results for $\eta$. We find that for $n=0$, $P(s)$ is close to $0.8$, while it fluctuates around $0.2$ for $n=1$. In other words, $P(s)$ roughly follows a Poisson distribution for $n=0$ and obeys the Wigner distribution for $n=1$. In this sense, this system changes from regular to chaotic. Normally, this transition is closely related to the system's symmetry\,\cite{Haake,ABmann2016}. Here, it is intimately connected with parity-symmetry-breaking in the thermodynamic limit $N\rightarrow\infty$ [ see Appendix \ref{app1}]. Specifically, due to the introduced optomechanical interaction, the proposed quantum model allows the single-photon-induced parity-symmetry-breaking under the condition of fixed system parameters.

\subsection{Sensitive dependence on perturbations} The hypersensitivity to perturbations is another signature of chaos\,\cite{Peres1984}. We will now calculate the Loschmidt Echo (LE) to quantify this sensitivity. The LE was first introduced in NMR experiments to measure the sensitivity to perturbations brought by the surrounding environment\,\cite{Pastawski2000,Usaj,Levstein,Karkuszewski2002,Jalabert2001,Quan2006}. The validity of this measurement has been verified experimentally in many-body spin systems. Quantum mechanics without the concept of trajectory preserves the overlap between two states. Specifically, for a quantum system, given the same initial state $\Psi_0$, under the influence of two Hamiltonians with a slight perturbation, the two states will evolve along with time\,\cite{Cucchietti2003}. More precisely, here we consider an extra two-level atom $S$ added into this hybrid optomechanical model\,\cite{Huang2009} as a perturbation. We assume that the hybrid DM is prepared in the ground state $|G\rangle=|00\rangle$ and the extra two-level atom is in a superposed state $\alpha|v\rangle+\beta|u\rangle$, where $|\alpha|^2+|\beta|^2=1$; accordingly the LE reads [see Appendix \ref{app2}]
\begin{align}
L(t)=|\langle G|{\rm exp}(iH_v t){\rm exp}(-iH_ut)|G \rangle|^2.
\end{align}
Here $H_v$ and $H_u$ are two Hamiltonians with a slight difference, $H_{v,u}=\omega_{v,u}b_n^{\dagger}b_n+\Omega d^{\dagger}d+\lambda_n(b_n^{\dagger}+b_n)(d^{\dagger}\sqrt{1-d^{\dagger}d/N}+{\rm H.c.})$, where $\omega_v=\omega_n+\tilde{\delta}$, $\omega_{u}=\omega_n-\tilde{\delta}$, and $\tilde{\delta}$ is a small perturbation caused by the extra two-level atom. Here we have applied the Holstein-Primakoff transformation: $J_{+}=d^{\dagger}\sqrt{N-d^{\dagger}d}, J_{-}=\sqrt{N-d^{\dagger}d}\,d$, and $J_{z}=d^{\dagger}d-N/2$, where $d$ is the bosonic operator\,\cite{Holstein1949}. In quantum mechanics, the overlap between the two identical initial states is supposed to be 1. Then it decays along with the evolution of the two states under the influence of two Hamiltonians. In some cases, after a time evolution, if these two wavefunctions become quite different, or completely different (orthogonal), then $L(t)$ should be much less than 1 or equal to 0. This shows that the system exhibits a hypersensitive dependence on the initial perturbation, i.e., chaotic behavior.
\begin{figure}
\centerline{\includegraphics[width=8.9cm]{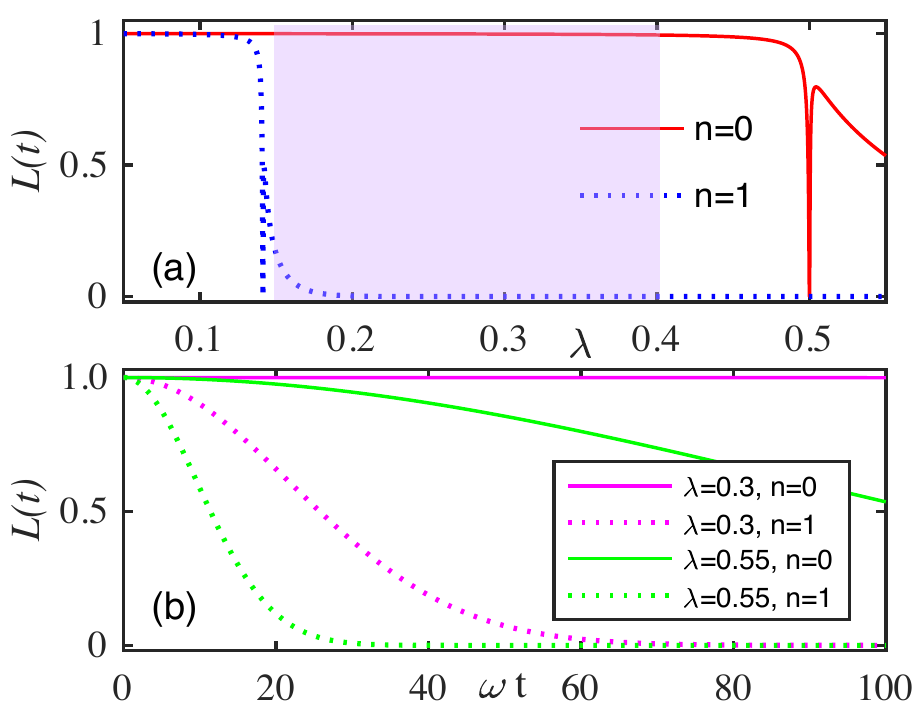}}
\caption{(a) The $L(t)$ in Eq.\,(3) versus the qubit-field coupling strength $\lambda$. We color the region where single-photon triggered chaos could be observed. (b) Evolutions of LE for various values of $\lambda$. We choose $\tilde{\delta}=0.001\omega$ and $\omega t=100$ for (a) and $N=100$ for both.}
\label{fig3}
\end{figure}
\begin{figure*}
\centerline{\includegraphics[width=16.6cm]{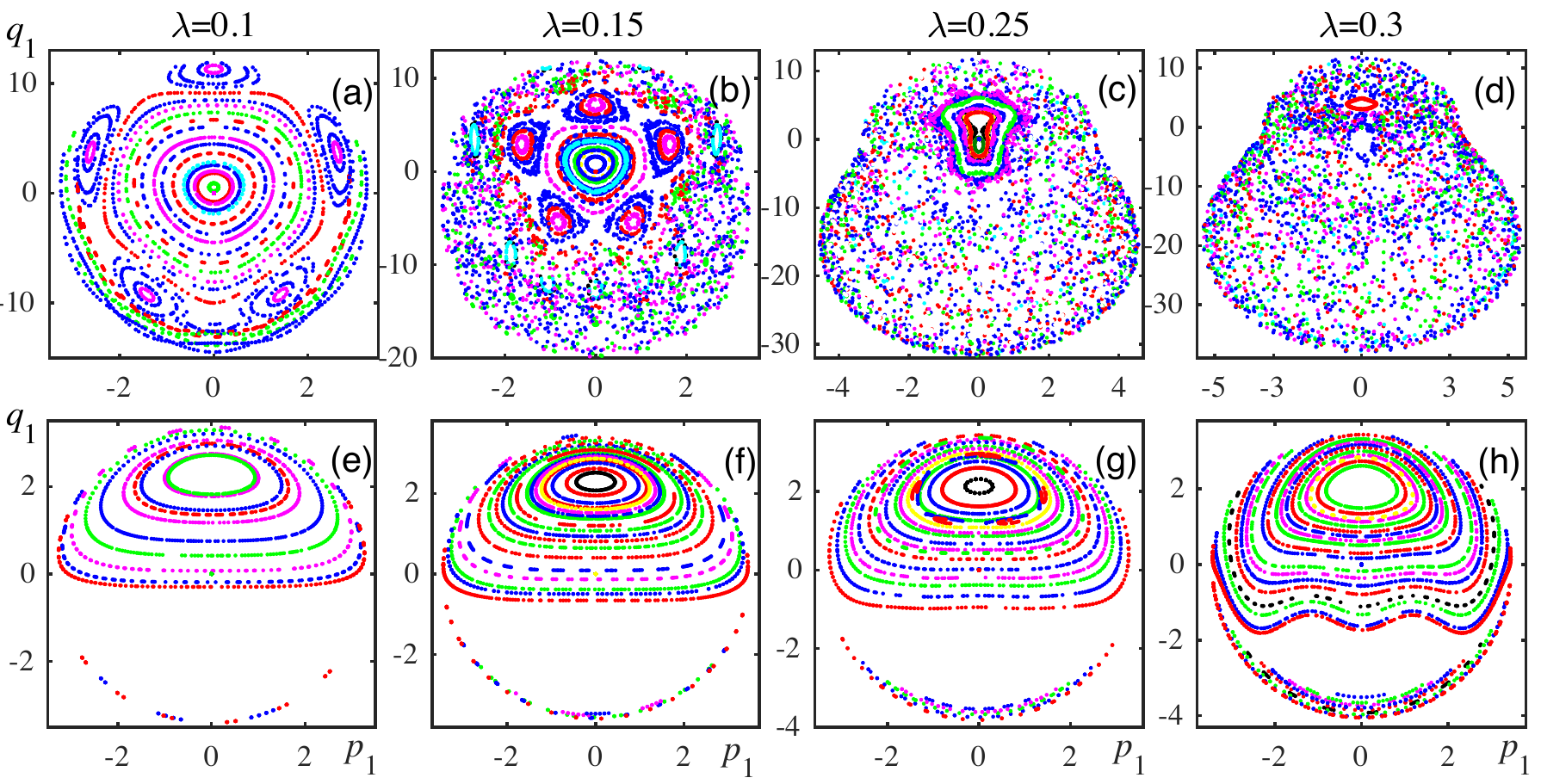}}
\caption{Plots of Poincar\'{e} surface sections projected in the $p_1$--$q_1$ plane for the Hamiltonian $H_{\rm sc}$ for various values of $\lambda$. We select sections of $E=-\omega$, and $p_2=0$, $q_2>0$. The top row (a)-(d) show $n=1$ and the bottom (e)-(h) present $n=0$. Each column has the same coupling strength $\lambda$.}
\label{fig4}
\end{figure*}

Fig.\,\ref{fig3}(a) plots the LE versus the coupling strength $\lambda$, showing that the LE decays quickly to zero when the coupling approaches a critical point $\lambda_{nc}=\sqrt{\omega_n\Omega}/2$. This proves that when the qubit-field coupling is near this critical point, the proposed system is very sensitive to the perturbation caused by the extra atom. We find that, compared to the case of $n=0$, $L(t)$ decreases more sharply at a lower $\lambda$ with $n=1$. Thus, this system exhibits a stronger sensitivity under the influence of an extra atom, compared with the case without optomechanical interactions ($n=0$). In essence, the optomechanical interaction lowers the critical point. When $\lambda\in [0.15,0.4]$, there is a high probability for the emergence of quantum chaos triggered by a single photon (see the shaded area). Figure\,\ref{fig3}(b) plots the evolution of LE for various values of $\lambda$. For a given $\lambda$, $L(t)$ for $n=1$ decreases faster than the case of $n=0$, which indicates that the two same initial states at $n=1$ evolve faster (than the $n=0$ case) to entirely different states after a long enough time. The corresponding arrows also show the emergence of single-photon-triggered quantum chaos.

\subsection{Poincar\'{e} Sections} In analogy with classical chaos, we proceed to consider the classical counterpart of the proposed system. By introducing the bosonic modes for position-momentum representation via $b_n=\sqrt{\omega_n/2}(q_1+ip_1/\omega_n), d=\sqrt{\Omega/2}(q_2+ip_2/\Omega)$, and setting $[q_1,p_1]=0$ and $[q_2,p_2]=0$, we could move from the Hamiltonian $H_n$ to a classical one,
\begin{align}
 H_{\rm cl}=&-\frac{N}{2}\Omega+\frac{1}{2}(\omega_n^2q_1^2+p_1^2-\omega_n+\Omega^2q_2^2+p_2^2-\Omega)  \nonumber \\
 &+2\lambda_n\sqrt{\Omega\omega_n}q_1p_1\sqrt{1-\frac{\Omega^2q_2^2+p_2^2-\Omega}{2N\Omega}}.
\end{align}
We calculate the derivative of $H_{\rm cl}$, and then obtain the equations of motion [see Appendix \ref{app3}],
\begin{align}
\!\!\!\!\dot{q_1}=\frac{{\rm \partial} H_{\rm cl}}{{\rm \partial} p_1}, \,\,
\dot{p_1}=-\frac{{\rm \partial} H_{\rm cl}}{{\rm \partial} q_1},  \,\,
\dot{q_2}=\frac{{\rm \partial} H_{\rm cl}}{{{\rm \partial}} p_2},  \,\,
\dot{p_2}=-\frac{{\rm \partial } H_{\rm cl}}{{\rm \partial}q_2}. \label{de}
\end{align}

To better visualize the presence of single-photon triggered chaos in the classical regime, we now study the dynamics of the canonical variables as a function of the coupling parameter $\lambda$. Specifically, we numerically integrate this simultaneous classical equations of motion and plot the Poincar\'{e} sections, which have been experimentally observed in many systems\,\cite{Lemos2012,Steck2001}, for a variety of different initial conditions. The resulting Poincar\'{e} surface of sections with energy $E=-\omega$ are shown in Fig.\,\ref{fig4}. When $n=0$, i.e., Figs.~(e-h), the Poincar\'{e} section only consists of a series of regular trajectories and discrete islands in the region considered. For the case $n=1$ shown in Figs.~(a-d), as the coupling strength $\lambda$ is increased, we see the shrinking of regular orbits and the expansion of a series of chaotic regions. For a certain Poincar\'{e} section, the motion across the boundaries between regular and chaotic regions is classically forbidden\,\cite{Cucchietti2003}. Interestingly, for a fixed coupling strength $\lambda$, by comparing the  bottom row for $n=0$ with the top for $n=1$, it is clearly shown that the classical trajectories transform from a regular to a chaotic one in phase space when the parameter $\lambda\in[0.15,0.4]$. In other words, it manifests single-photon-triggered chaos. The observed Poincar\'{e} sections show an excellent agreement with the prediction of nearest-level distribution (shown in Fig.\,\ref{fig2}) and the dynamics properties (shown in Fig.\,\ref{fig3}).

\section{Experimental implementations} 

The description of the quantized electromagnetic field is based on states with fixed photon number in specific modes, i.e., Fock or number states. Recent years, there has been increasing interest in the generation of single-photon states\,\cite{Hong1986,Lvovshy2001,Varcoe2000,Brattke,You2007,Maitre,Hofheinz2008,Hofheinz2009,Liu2004,Chu2018,Gu2017} because single photons play a key role in secure quantum communication, quantum cryptography, and various single-photon devices. Single-photon state is easily prepared conditionally from photon-pairs of parametric down-conversion, ion traps, cavity QED systems, and solid-state systems using superconducting quantum circuits.

In the experiment\,\cite{Varcoe2000}, a micromaser (or one-atom maser) employed a high--$Q$ ($4\times 10^{10}$) cavity producing a photon lifetime in the cavity as high as 0.3 seconds. This system can be described by the Jaynes-Cummings Hamiltonian. Traditionally, the injected atoms are in the upper state of the maser transition. If the field is in an initial state $|n\rangle$, then the interaction of an atom with a field leaves the cavity and atom field in a superposition of the states $|e\rangle|n\rangle$ and $|g\rangle|n+1\rangle$. By measuring an atom in the ground state, the superposition is reduced to the state $|n+1\rangle$. Experimentally, it is easy to measure the atom inversion given by $I=P_g-P_e$; here $P_g$ and $P_e$ are the probability of finding a ground-state or excited-state atom, respectively.

Solid-state systems are considered as another good candidate for generating Fock states. Recently, an on-chip microwave single-photon source was reported in a circuit quantum electrodynamics system consisting of a superconducting qubit coupled to a transmission line\,\cite{Hofheinz2008}. It was reported that a six-photons pure Fock states can be prepared in this architecture. The interaction between qubit and resonator is achieved by using a capacitor, measured by spectroscopy. The qubit states are measured by using a read-out d.c. superconducting quantum interference device (SQUID). If the qubit state is mapped to the photon state, then the superposition of the ground and excited states $\alpha|g\rangle+\beta|e\rangle$, will lead to the same superposition of photon states: $\alpha|0\rangle+\beta|1\rangle$. The average photon number is proportional to the average qubit excitation probability $\langle a^{\dagger}a\rangle=(\langle \sigma_z\rangle+1)/2$, and has a maximum of one photon when
the qubit is in its excited state. The two quadratures of
homodyne voltage are proportional to the $x$ and
$y$ components of the qubit state: $\langle a^{\dagger}+a\rangle=\langle \sigma_x\rangle$ and $i\langle a^{\dagger}-a\rangle=\langle \sigma_y\rangle$. Experimentally, the agreements between output integrated voltage of states $\langle a^{\dagger}a\rangle$ and $\langle \sigma_z\rangle$, $i\langle a^{\dagger}-a\rangle$ and $\langle \sigma_y\rangle$, respectively, verify the generation of Fock states.

The proposed system is a general quantum hybrid model, which can be realized in several experimental platforms, e.g., quadratically-coupled optomechanical system or superconducting circuit. Thanks to the rapid progress achieved in the fabrication of diamond nanostructures, the Dicke model can be implemented in diamond mechanical nanoresonators\,\cite{Arcizet2011,Bennett2013,Teissier2014,Pengbo2016}, i.e., an ensemble of nitrogen-vacancy (NV) centers embedded in a single crystal diamond nanobeam. The flex of the beam strains the diamond lattice and, in turn, couples directly to the spin triplet states; then a crystal-strain-induced coupling can be generated. Moreover, the required quadratic coupling can be realized in a ``membrane-in-the-middle" configuration, by placing a semitransparent membrane at the node of the ancillary cavity mode, where $\omega^{'}=0$\,\cite{Thompson2008,Bhattacharya2008}.

In addition, as an alternative experimental possibility, a hybrid quantum superconducting circuit \,\cite{Pengbo2018,Kakuyanagi2016,Kubo2011} is also considered here. In the superconducting circuit, the qubits are coherently coupled to the spin ensemble, which can be seen as a two-level atom ensemble. The Dicke model can be realized by capacitively coupling the qubits to the resonator B\,\cite{Gui-Lei20181,Gui-Lei20182}. Moreover, resonator A contains superconducting quantum-interference devices (SQUIDs) that makes its frequency $\omega_a$ tunable by applying opposite flux variations $\pm\delta\Phi$ in its loop. Due to this parametrically-induced frequency shift of resonator A, the position quadrature of resonator B can couple quadratically to the photon number of resonator A in a certain regime\,\cite{Kim2015}.

To implement our proposal, the key is to reach the strong quadratic coupling $g\approx\omega/4$ ($g\approx\omega_c^{''}(0)x_{\rm zpf}^2$). Fortunately, recent advances have shown that stronger quadratic couplings could be achieved in many platforms, such as, photonic crystals\,\cite{Kalaee2016,Paraiso2015}, superconducting circuits\,\cite{Heikkila2014} and microdisk resonators\,\cite{Doolin2014}. In addition, It is manifested that the quadratic coupling strength could be exponentially enhanced \cite{Lu2015,Taishang2017}, which allows the realization of single-photon-triggered quantum chaos when $g\ll \omega$.

{\color{black}\section{Conclusions} In summary, we have studied quantum chaos triggered by a single photon in a cavity optomechanical system doped with an atom ensemble. By preparing different states of the ancillary mode, the system can be changed between quasi-integrable regime and quantum chaos. This novel but fundamental effect could be useful in engineering new types of quantum on-chip devices, simulating various single-photon devices and achieving chaos-assisted communications.}

\begin{acknowledgments} We are very grateful to P. P.-Fern\'{a}ndez, M. Vyas and T. Scholak for valuable discussions. X.Y.L and Y.W are supported by the National Key Research and Development Program of China grant 2016YFA0301203, the National Natural Science Foundation of China (Grant Nos. 11822502, 11374116, 11574104 and 11375067). F.N. is supported in part by the: MURI Center for Dynamic Magneto-Optics via the Air Force Office of Scientific Research (AFOSR) (FA9550-14-1-0040), Army Research Office (ARO) (Grant No. W911NF-18-1-0358), Asian Office of Aerospace Research and Development (AOARD) (Grant No. FA2386-18-1-4045),
Japan Science and Technology Agency (JST) (via the Q-LEAP program, and the CREST Grant No. JPMJCR1676),
Japan Society for the Promotion of Science (JSPS) (JSPS-RFBR Grant No. 17-52-50023, and JSPS-FWO Grant No. VS. 059. 18N),
the RIKEN-AIST Challenge Research Fund, and the John Templeton Foundation.
\end{acknowledgments}

\appendix
\section{Single-photon-induced ${\mathbb Z_2}$ symmetry breaking}\label{app1}
In the main text, we obtained the Dicke-like Hamiltonian,

\begin{align}
H_{n}=\Omega J_{z}+\omega_n b_{n}^{\dagger}b_n+\frac{\lambda_n}{\sqrt{N}}(b_{n}^{\dagger}+b_n)J_{x}+C_n,\label{H_dn}
\end{align}
where
$\omega_n={\rm exp}(-2r_n)\omega, $
$\lambda_n={\rm exp}(r_n)\lambda,$
 and $
 C_n=\sum_{l=e,o}n_l\omega_l+[{\rm exp}(-2r_n)-1](\omega/2),$
 and correspondingly, the critical coupling strength reads
 \begin{align}\lambda_{nc}=\sqrt{\Omega\omega_n}/2.\end{align} The Hamiltonian(\ref{H_dn}) shows a photon-dependent property of our proposed system. This property makes it possible to observe single-photon-triggered quantum chaos, which is closely connected with the single-photon-induced ${\mathbb Z_2}$-symmetry-breaking in the thermodynamic limit $N\rightarrow\infty$. To investigate the system's symmetry, as an example, we consider a certain qubit-field coupling strength $\lambda=0.3$ (within our considered region $0.15<\lambda<0.4$) and the cases when $n=0,1$, respectively.

In the case when $n=0$, Equation\,(\ref{H_dn}) can be reduced into a normal Dicke Hamiltonian.  The system's critical point becomes $\lambda_c=\sqrt{\Omega\omega}/2=0.5$, and here we have chosen $\Omega=\omega=1$. Within the considered parameter regime of $\lambda<\lambda_c$, the system is in the normal phase. The present effective Hamiltonian $H_n$ could be diagonalized in the thermodynamic limit $N\rightarrow\infty$\,\cite{Emary2003}.
To diagonalize the Hamiltonian $H_{n}$, we introduce the Holstein-Primakoff transformation, $J_{+}=d^{\dagger}\sqrt{N-d^{\dagger}d}, J_{-}=\sqrt{N-d^{\dagger}d}\,d$, and $J_{z}=d^{\dagger}d-N/2$\,\cite{Bhattacharya2008}. Then Eq.\,(\ref{H_dn}) can be diagonalized into
\begin{align}
H_{\rm np}=\omega_{-}c_{1}^{\dagger}c_{1}+\omega_{+}c_{2}^{\dagger}c_2+E_g,\label{H_np}
\end{align}
here
\begin{align}\omega^{2}_{\pm}=\frac{1}{2}\left[\omega_n^2+\Omega^2\pm\sqrt{(\Omega^2-\omega_n^2)^2+16\lambda_n^2\Omega\omega_n}\right].
\end{align}
We have introduced the Bogoliubov transformation,
\begin{subequations}
\begin{align}
b_n=\,&\xi^{(b)}_{-}c^{\dagger}_{1}+\xi^{(b)}_{+}c_{1}+\zeta^{(b)}_{-}c^{\dagger}_{2}+\zeta^{(b)}_{+}c_{2},
\\
d=\,&\xi^{(d)}_{-}c^{\dagger}_{1}+\xi^{(d)}_{+}c_{1}+\zeta^{(d)}_{-}c^{\dagger}_{2}+\zeta^{(d)}_{+}c_{2},
\end{align}\label{bd}
\end{subequations}
where the coefficients satisfy

\begin{align}
\xi^{(b)}_{\pm}=&\frac{\cos\nu}{2\sqrt{\omega_n\omega_{-}}}(\omega_n\pm\omega_{-}),\,\,\,\,\,\,\,\,\zeta^{(b)}_{\pm}=\frac{\sin\nu}{2\sqrt{\omega_n\omega_{+}}}(\omega_n\pm\omega_{+}),
\nonumber\\
\xi^{(d)}_{\pm}=&-\frac{\sin\nu}{2\sqrt{\Omega\omega_{-}}}(\Omega\pm\omega_{-}), \,\,\,\,\,\,\,\,\zeta^{(d)}_{\pm}=\frac{\cos\nu}{2\sqrt{\Omega\omega_{+}}}(\Omega\pm\omega_{+}).\label{xizeta}
\end{align}
Here the angle $\nu$ is determined by \begin{align}\tan(2\nu)=\frac{4\lambda_n\sqrt{\Omega\omega_n}}{(\Omega^2-\omega^2_n)}.\end{align}
In this case, the ground state reads $|G\rangle_{\rm np}=|00\rangle_c$ and correspondingly, the ground-state energy is $E_g/N=-\Omega/2$\,\cite{Emary2003}. The ground state $|G\rangle_{\rm np}$ conserves the ${\mathbb Z}_2$ symmetry, i.e., $ \Pi|00\rangle_c=|00\rangle_c$, confirmed by the zero ground-state coherence of the field, i.e., $\langle b\rangle_g=0$.

In the other case, if we consider $n=1$, the effective coupling strength is enhanced and the critical qubit-field coupling strength is effectively decreased, i.e., $\lambda_{n}>\lambda_{nc}$. In other words, when $n=1$ and $\lambda=0.3$, the system reaches into the superradiant phase. By applying two displacements on the bosonic modes
\begin{align}
b_n\rightarrow\tilde{b}_n+\gamma_b, \,\,\,\,\,d\rightarrow\tilde{d}-\gamma_d
\end{align}
or
\begin{align}
b_n\rightarrow\tilde{b}_n-\gamma_b, \,\,\,\,\,d\rightarrow\tilde{d}+\gamma_d,
\end{align}
and in the thermodynamic limit, the Hamiltonian $H_{ n}$ can be diagonalized to
\begin{align}H_{\rm sp}=\tilde{\omega}_{-}\tilde{c}_{1}^{\dagger}\tilde{c}_{1}+\tilde{\omega}_{+}\tilde{c}_{2}^{\dagger}\tilde{c}_{2}+\tilde{E}_{g}.
\end{align}
We note that
\begin{align}\gamma_b=\sqrt{N(\frac{\lambda_n^2}{\omega_n^2}-\frac{\Omega^2}{16\lambda_n^2})},\,\,\,\,\,\,{\rm and} \,\,\,\,\,\gamma_d=\sqrt{\frac{N}{2}(1-\frac{\Omega\omega_n}{4\lambda_n^2})}.\end{align}
Following the same process as before, but now using
\begin{subequations}
\begin{align}
\tilde{b}_n=\,\,&\tilde{\xi}^{(b)}_{-}\tilde{c}^{\dagger}_{1}+\tilde{\xi}^{(b)}_{+}\tilde{c}_{1}+\tilde{\zeta}^{(b)}_{-}\tilde{c}^{\dagger}_{2}+\tilde{\zeta}^{(b)}_{+}\tilde{c}_{2},
\\
\tilde{d}=\,\,&\tilde{\xi}^{(d)}_{-}\tilde{c}^{\dagger}_{1}+\tilde{\xi}^{(d)}_{+}\tilde{c}_{1}+\tilde{\zeta}^{(d)}_{-}\tilde{c}^{\dagger}_{2}+\tilde{\zeta}^{(d)}_{+}\tilde{c}_{2},
\end{align}\label{bd2}
\end{subequations}
the coefficients are then given by

\begin{align}
\tilde{\xi}^{(b)}_{\pm}=&\frac{\cos\tilde{\nu}}{2\sqrt{\omega_n\tilde{\omega}_{-}}}(\omega_n\pm\tilde{\omega}_{-}), \,\,\,\,\,\,\, \tilde\zeta^{(b)}_{\pm}=\frac{\sin\tilde{\nu}}{2\sqrt{\omega_n\tilde{\omega}_{+}}}(\omega_n\pm\tilde{\omega}_{+}),
\nonumber\\
\tilde{\xi}^{(d)}_{\pm}=&-\frac{\sin\tilde{\nu}}{2\sqrt{\tilde{\Omega}\tilde{\omega}_{-}}}(\tilde{\Omega}\pm\tilde{\omega}_{-}), \,\,\,\,\,\,\, \tilde{\zeta}^{(d)}_{\pm}=\frac{\cos\tilde{\nu}}{2\sqrt{\tilde{\Omega}\tilde{\omega}_{+}}}(\tilde{\Omega}\pm\tilde{\omega}_{+}).\label{xizeta2}
\end{align}
Here the angle $\tilde{\nu}$ satisfies \begin{align}\tan(2\tilde{\nu})=\frac{2\omega_n\Omega}{(16\lambda_n^4/\omega_n^2-\omega^2_n)}\end{align} and $\tilde{\Omega}=\Omega(1+\frac{4\lambda_n^2}{\Omega\omega_n})/2$.
The excitation energies become
\begin{align}
\tilde{\omega}^2_{\pm}=\frac{1}{2}\left[\omega^2_n+16\lambda_n^4/\omega_n^2\pm\sqrt{(16\lambda_n^4/\omega_n^2-\omega_n^2)^2+4\Omega^2\omega^2_n}\right].\end{align}
We point out that the ground-state energy is
\begin{align}
\tilde{E}_g=-\frac{\Omega}{4}(\frac{4\lambda_n^2}{\Omega\omega_n}+\frac{\Omega\omega_n}{4\lambda_n^2})\end{align}
and the ground state $|G\rangle_{\rm sp}^{\pm}=|00\rangle_c^{\pm}$ with $\tilde{c}_i^{\dagger}\tilde{c}_i|00\rangle_c^{\pm}=0|00\rangle_c^{\pm} (i=1,2)$ becomes two-fold degenerated. Here $\pm$ represent the different direction of displacement applied into $b_n$, which leads to different coefficients $\tilde{\xi}_{\pm}$ and $\tilde{\zeta}_{\pm}$. As an evidence, the ground-state coherence of the field $\langle b\rangle_g^{\pm}=\pm {\rm exp}(r_n)\gamma_b$ becomes nonzero. Consequently, its symmetry is spontaneously broken, i.e, $\Pi|G\rangle_{\rm sp}^{\pm}\neq|G\rangle_{\rm sp}^{\pm}$.

The above discussions show the single-photon-induced $\mathbb Z_2$-symmetry-breaking in the thermodynamic limit $N\rightarrow \infty$. Specifically, for a fixed qubit-field coupling $\lambda$ (within our considered region), the system's symmetry can be broken by injecting a single photon. This symmetry breaking is closely related to the crossover of the level statistics from a Poisson-like distribution to a Wigner-like form\,\cite{ABmann2016,Haake} for finite $N$. Accordingly, it is connected to the occurrence of quantum chaos.

\section{derivation of Loschmidt Echo}\label{app2}

In order to illustrate how chaos is characterized by the hypersensitivity to perturbations, we analytically calculate the Loschmidt Echo (LE) of the proposed system. For a quantum system, the measure of the LE is the overlap between two states that evolve from the same initial state $\Psi_{0}$ under the influence of two Hamiltonians. Specifically, here we consider an extra two-level atom $S$ injected into this hybrid quantum model\,\cite{Huang2009}. Then the system Hamiltonian is given by (we set $\hbar=1$),
\begin{subequations}
\begin{align}
H_0&=\sum_{l=e,o}\omega_l a_l^{\dagger}a_l+\omega b^{\dagger}b+\Omega J_z+\frac{\lambda}{\sqrt{N}}(b^{\dagger}+b)(J_+ +J_-)\nonumber\\&-\sum_{l=e,o}g_la_l^{\dagger}a_l(b+b^{\dagger})^2,   \\
H_I&=\frac{1}{2}\omega_s\sigma_z+\lambda_s(b^{\dagger}+b)(\sigma_+ +\sigma_-).\label{H_0i}
\end{align}
\end{subequations}
Here the Hamiltonian $H_{I}$ describes the extra two-level atom $S$ with transition operators $\sigma_z$, $\sigma_+$ and $\sigma_{-}$,  coupled to the single-mode bosonic field of the initial normal Dicke model.  Also, $\omega_s$ is the frequency between the ground state $|v\rangle$ and excited state $|u\rangle$ of the extra atom $S$. Same as before, we prepare the $a_l$ mode in the Fock state $|n_e,n_o\rangle_a$ $(n_e=n_o+n$, and $n_e,n_o,n=0,1,2...)$ and apply a squeezing transformation $b=\cosh(r_{n})b_{n}+\sinh(r_{n})b_{n}^{\dagger}$. Provided that the detuning between atom $S$ and the single-mode bosonic mode $b$ is much larger than the coupling strength $\lambda_s$, i.e., $|\Delta_s|\equiv|\omega_s-\omega|\gg\lambda_s$, applying the Fr\"{o}hlich-Nakajima transformation, then we obtain
\begin{align}
H_{\rm eff}&=(\omega_n+\tilde{\delta}\sigma_z)b_{n}^{\dagger}b_n+\frac{1}{2}(\omega_s+\tilde{\delta})\sigma_z+\Omega J_z\nonumber\\&+\frac{\lambda_n}{\sqrt{N}}(b_{n}^{\dagger}+b_n)J_{x}+C_n,\label{H_eff}
\end{align}
where $\tilde{\delta}=\lambda_{s}^2/\Delta_s$, and we have made a rotating wave approximation. We note that the second term in $H_{\rm eff}$ describes the extra two-level atom. As before, we use the Holstein-Primakoff transformation, the Hamiltonian $H_{\rm eff}$ is further reduced to
\begin{align}
H_{\rm eff}&=(\omega_n+\tilde{\delta}\sigma_z)b_{n}^{\dagger}b_n+\frac{1}{2}(\omega_s+\tilde{\delta})\sigma_z+\Omega d^{\dagger}d\nonumber\\&+\lambda_n(b^{\dagger}_n+b_n)\left( d^{\dagger}\sqrt{1-\frac{d^{\dagger}d}{N}}+\sqrt{1-\frac{d^{\dagger}d}{N}}d\right)+C_n.   \label{H_eff}
\end{align}

We assume that the extra atom $S$ is in a superposed state $\alpha|v\rangle+\beta|u\rangle$, here $|\alpha|^2+|\beta|^2=1$. Then Eq.\,(\ref{H_eff}) could be rewritten as,
\begin{align}
H_{\rm eff}=|v\rangle\langle v|\otimes H_v+|u\rangle\langle u|\otimes H_u,
\end{align}
with
\begin{align}
H_v&=\omega_v b_{n}^{\dagger}b_n+\Omega(d^{\dagger}d-N/2)\nonumber \\&+\lambda_n(b^{\dagger}_n+b_n)\left( d^{\dagger}\sqrt{1-\frac{d^{\dagger}d}{N}}+\sqrt{1-\frac{d^{\dagger}d}{N}}d\right)+C_n, \label{H_v}
\end{align}
and
\begin{align}
H_u&=\omega_u b_{n}^{\dagger}b_n+\Omega(d^{\dagger}d-N/2)\nonumber\\&+\lambda_n(b^{\dagger}_n+b_n)\left( d^{\dagger}\sqrt{1-\frac{d^{\dagger}d}{N}}+\sqrt{1-\frac{d^{\dagger}d}{N}}d\right)+C_n,\label{H_e}
\end{align}
where $\omega_v=\omega_n-\tilde{\delta}$ and $\omega_u=\omega_n+\tilde{\delta}$. By comparing $H_{v} (H_{u})$ with the Hamiltonian $H_n$, we find that the extra two-level atom only changes the frequency of the effective single-mode bosonic mode with $\tilde{\delta}$. At time $t$, the total state becomes,
\begin{align}
|\Psi (t)\rangle&=e^{-iH_{\rm eff}t}(\alpha |v\rangle+\beta|u \rangle)\otimes|G\rangle,  \nonumber \\
&=\alpha|v\rangle\otimes e^{-iH_{v}t}|G\rangle+\beta|u\rangle\otimes e^{-iH_{u}t}|G\rangle. \label{Psi}
\end{align}
We have assumed that the photon-dressed atomic ensemble is initially in the ground state $|G\rangle=|00\rangle$ of the Hamiltonian $H_n$.
In the following, we will prove that the dynamics of the photon-dressed atomic ensemble is sensitive to the state of the extra atom. We trace over the degrees of freedom of the photon-dressed atomic ensemble in $|\Psi (t)\rangle$, then the reduced density matrix reads
\begin{align}
\rho(t)=|\alpha|^2|v\rangle\langle v|+|\beta|^2|u\rangle\langle u|+(D\alpha ^*\beta|u\rangle\langle v|+{\rm H.c.}),\label{rho}
\end{align}
here $D$ is a decoherence factor, which is given by
\begin{align}
D(t)=\langle G|{\rm exp}(iH_v t){\rm exp}(-iH_u t)|G\rangle.\label{D}
\end{align}
The Loschmidt Echo can be defined as
\begin{align}
L(t)=|D(t)|^2,
\end{align}
and thus we obtain Eq.\,(3) in the main text. For a short time $t$, $L(t)$ could be approximated as
\begin{align}
L(t)\approx {\rm exp}(-4\varrho\tilde{\delta}^2 t^2)\label{Lt}
\end{align}
with $\varrho=\langle G|(b_n^{\dagger}b_n)^2|G\rangle-\langle G| b_{n}^{\dagger}b_n|G\rangle^2$. In the normal phase, we apply Eqs.\,(\ref{bd}) into Eq.\,(\ref{Lt}), and then we obtain
\begin{align}
\varrho_{\rm np}=2\xi^{(b)2}_{+}\xi^{(b)2}_{-}+2\zeta^{(d)2}_{+}\zeta^{(d)2}_{-}+(\xi^{(b)}_{+}\zeta^{(d)}_{-}+\xi^{(b)}_{-}\zeta^{(d)}_{+})^2.
\end{align}
In the superradiant phase,
we apply Eqs.\,(\ref{bd2}) into Eq.\,(\ref{Lt})\,(using $\tilde{b}_n$ to replace $b_n$), and then we obtain
\begin{align}
\varrho_{\rm sp}&=2\tilde{\xi}^{(b)2}_{+}\tilde{\xi}^{(b)2}_{-}+2\tilde{\zeta}^{(d)2}_{+}\tilde{\zeta}^{(d)2}_{-}+
(\tilde{\xi}^{(b)}_{+}\tilde{\zeta}^{(d)}_{-}+\tilde{\xi}^{(b)}_{-}\tilde{\zeta}^{(d)}_{+})^2\nonumber\\&+\gamma_b^2\left[(\tilde{\xi}^{(b)}_{+}
+\tilde{\xi}^{(b)}_{-})^2+(\tilde{\zeta}^{(d)}_{+}+\tilde{\zeta}^{(d)}_{-})^2\right].
\end{align}
In addition, we could define the purity $P={\rm Tr}_{S}(\rho_{S}^2)={\rm Tr}_{S}\{[{\rm Tr}_{E}\rho(t)]^2\}$\,\cite{Quan2006} to describe the evolution of two states. Here, $\rho(t)=|\Psi (t)\rangle\langle\Psi (t)|$, and Tr means tracing over the variables of $E$ or $S$. Then the purity $P$ can be written as
\begin{align}
P=1-2|\alpha\beta|^2[1-L(t)].\label{eq.P}
\end{align}
The purity demonstrates the degree of similarity between two states, which can test the system's sensitivity. Physically, the purity satisfies $P\le 1$. For two identical initial states, $P$ is supposed to be 1. High sensitivity to the perturbation should be reflected in a large decay of the purity.
\section{classical model}\label{app3}
In the following, we study the classical counterpart
of the proposed system. We move from the quantum mechanical Hamiltonian Eq.\,(\ref{H_dn}) to a classical one by introducing
\begin{align}
b_n&=\sqrt{\frac{\omega_n}{2}}\left(q_1+\frac{i}{\omega_n}p_1\right), \,\,\,\,\,\,\, b_n^{\dagger}=\sqrt{\frac{\omega_n}{2}}\left(q_1-\frac{i}{\omega_n}p_1\right),   \nonumber \\
d&=\sqrt{\frac{\Omega}{2}}\left(q_2+\frac{i}{\Omega}p_2\right),\,\,\,\,\,\,\,\,\,\,\,\,d^{\dagger}=\sqrt{\frac{\Omega}{2}}\left(q_2-\frac{i}{\Omega}p_2\right),
\end{align}
where $q_i,p_i (i=1,2)$ are the position and momentum operators, respectively. This Hamiltonian in the position-momentum representation can be written as
\begin{align}
&H_n^{'}=-\frac{N}{2}\Omega+\frac{1}{2}(\omega_n^2q_1^2+p_1^2-\omega_n+\Omega^2q_2^2+p_2^2-\Omega)\nonumber\\&+
\lambda_n\sqrt{\Omega\omega_n}q_1\left[(q_2-\frac{i}{\Omega}p_2)\sqrt{1-\eta}+\sqrt{1-\eta}(q_2+\frac{i}{\Omega}p_2)\right],
\end{align}
where $\eta$ satisfies

\begin{align}
\eta=\frac{\Omega^2 q_2^2+p_2^2-\Omega}{2N\Omega}\leq1.   \nonumber
\end{align}
We set $[q_1,p_1]=[q_2,p_2]=0$, and in terms of classical variables we have
\begin{align}
 H_{\rm cl}&=-\frac{N}{2}\Omega+\frac{1}{2}(\omega_n^2q_1^2+p_1^2-\omega_n+\Omega^2q_2^2+p_2^2-\Omega)
 \nonumber\\&+2\lambda_n\sqrt{\Omega\omega_n}q_1p_1\sqrt{1-\frac{\Omega^2q_2^2+p_2^2-\Omega}{2N\Omega}}.
\end{align}
To analyze the behavior of this classical system for finite $N$, we derive the Hamiltonian $H_{\rm cl}$,
\begin{subequations}
\begin{align}
\frac{{\rm d}q_1}{{\rm d}t}&=\frac{{\rm\partial} H_{\rm cl}}{{\rm\partial} p_1}=p_1, \,\,  \\
\frac{{\rm d}p_1}{{\rm d}t}&=-\frac{\partial H_{\rm cl}}{\partial q_1}=p_2(1-\frac{\lambda_n}{N}\sqrt{\frac{\omega_n}{\Omega}}\frac{xy}{\sqrt{1-\eta}}),  \,\,\\
\frac{{\rm d} q_2}{{\rm d}t}&=\frac{\partial H_{\rm cl}}{\partial p_2}=-\omega_{n}^2x-2\lambda_n\sqrt{\Omega\omega_{n}}y\sqrt{1-\eta},  \,\,\\
\frac{{\rm d} p_2}{{\rm d}t}&=-\frac{\partial H_{\rm cl}}{\partial q_2}=-\Omega^2y-2\lambda_n\sqrt{\Omega\omega_n}x\sqrt{1-\eta}(1-\frac{\Omega y^2}{2N(1-\eta)}).
\end{align}
\end{subequations}

\end{document}